\documentclass[sigconf,authorversion]{acmart}

\usepackage[oldenum]{paralist}
\usepackage{booktabs}
\usepackage{amsmath}
\usepackage{amssymb}
\usepackage{verbatim}
\usepackage{subfigure}
\usepackage{algorithm}
\usepackage{algorithmic}
\usepackage{tikz}
\usepackage{pgfplots}
\usepackage{array}
\usepackage{multirow}
\usepackage{enumerate}
\usepackage[show]{chato-notes}

\renewcommand{\paragraph}[1]{\medskip\noindent\textbf{#1.~}}

\newcommand{\iadh}[1]{\textcolor{black}{#1}}
\newcommand{\zaiq}[1]{\textcolor{black}{#1}}
\newcommand{\zaiqm}[1]{\textcolor{black}{#1}}
\newcommand{\ric}[1]{\textcolor{black}{#1}}
\newcommand{\ricB}[1]{\textcolor{black}{#1}}

\newcommand*{\mat}[1]{\mathbf{#1}}
\newcommand*{\set}[1]{\mathcal{#1}}

\def \ours{VBCAR}
\def \tab{Table}
\def \fig{Figure}
\def \eq{Equation}
\def \sec{Section}

\copyrightyear{2019}
\acmYear{2019}
\setcopyright{rightsretained}
\acmConference[ACM RecSys]{ACM RecSys}
\acmBooktitle{Workshop on Context-Aware Recommender Systems}
\acmPrice{15.00}
\acmDOI{10.1145/xxxxxxx.xxxxxxx}
\acmISBN{}
\begin{document}

\title{Variational Bayesian Context-aware Representation for Grocery Recommendation}

\author{Zaiqiao Meng, Richard McCreadie, Craig Macdonald and Iadh Ounis}
\affiliation{\vspace{0.1cm}
\institution{University of Glasgow, Scotland, UK}
}
\email{(firstname.lastname)@glasgow.ac.uk}

\begin{abstract}
\ric{Grocery recommendation is an important recommendation use-case, which aims} to predict which items a user \iadh{might choose to buy} in the future\ric{, based on} their shopping history. However, \zaiqm{existing} methods only represent each user and item by single deterministic points in a low-dimensional continuous space. \iadh{In addition, most of these methods} are trained by maximizing the co-occurrence likelihood with a simple Skip-gram-based formulation, which limits the expressive ability of their embeddings and \zaiqm{the resulting} recommendation performance. In this paper, we propose \ric{the} Variational Bayesian Context-Aware Representation (\ours{}) model for grocery recommendation, which is a novel variational Bayesian model that learns \iadh{the} user and item latent vectors by leveraging basket context information from \ric{past} user-item interactions.
We train our \ours{} model based \iadh{on} the Bayesian Skip-gram framework couple\iadh{d} with the amortized variational inference, so that it can learn more expressive latent representations \ric{that integrate} both \zaiqm{the non-linearity and \iadh{Bayesian behaviour}}. Experiments conducted on \iadh{a large} \ric{real-world} grocery \ric{recommendation} dataset show that our \ric{proposed} \ours{} \iadh{model can} \ric{significantly} \iadh{outperform} \iadh{existing} state-of-the-art grocery recommendation methods.
\end{abstract}


\maketitle

\section{Introduction}
Recommender systems \ric{that use} historical customer-product interactions to provide customer\iadh{s} with useful suggestions \iadh{have} been of interest to \iadh{both} academia and industry for many years. Various matrix completion\iadh{-based} methods \cite{rendle2009bpr,he2017neural,mnih2008probabilistic} have been proposed to predict the rating scores of \zaiqm{products (or items)} for customers (or users). Recently, many grocery recommendation \ric{models}~\cite{wan2018representing,wan2017modeling,grbovic2015commerce} were proposed \ric{that} \iadh{target} grocery shopping \ric{use-cases}. \ric{In} real grocery shopping platforms, such as Amazon and Instacart, users' interactions \iadh{with} items are sequential, personalized and more complex than those represented by a single rating score matrix. Thus \ric{effective} recommendation models \ric{for this use-case} are designed to learn representations of users and items so that context\iadh{ual} information, such as basket context~\cite{wan2018representing} and time context \cite{manotumruksa2018contextual}, \ric{are} captured \ric{within the} learned representations, which \ric{results in increased} recommendation performance.
In the grocery shopping \ric{domain}, prod2vec~\cite{grbovic2015commerce} and triple2vec~\cite{wan2018representing} are two state-of-the-art models that learn latent representations capturing the basket context\ric{,} based on the Skip-gram model for grocery recommendation. In these models, both the user's general interest (which item\iadh{s} the user likes) and the personalized dependencies between \iadh{items} (what items the user \ricB{commonly includes} in the same basket) are encoded by the embeddings of users and items. \ric{Furthermore, when combined with} negative sampling \ric{approaches}~\cite{mikolov2013distributed}, these Skip-gram-based models are able to scale to \ric{very} large shopping datasets. \ric{Meanwhile, through the} \iadh{incorporation of} basket \ric{contextual information during representation learning}, \ricB{significant} \ric{improvements} in grocery recommendation \iadh{have been observed}~\cite{grbovic2015commerce,wan2018representing}. 

However, these representation models still have several defects: (1) they represent each user and item by single deterministic points in a low-dimensional continuous space, which limits the expressive ability of their embeddings and recommendation performance\iadh{s};
(2) their model\iadh{s} are simply trained by maximizing the likelihood of recovering the purchase history, which is a point estimate solution that \iadh{is} more sensitive to \ricB{outliers when training}~\cite{barkan2017bayesian}.

To alleviate the aforementioned problems, we propose a \emph{Variational Bayesian Context-Aware Representation} model, abbreviated as \emph{\ours{}}, which extends the existing Skip-gram based representation models for grocery recommendation in two directions. \ric{First,} it jointly models \iadh{the} representation of users and items in a Bayesian manner, \iadh{which} represents users and items as \ric{(Gaussian)} distributions and \ric{ensures} \iadh{that} these probabilistic representations \iadh{are} \ric{similar} to their prior distributions \ric{(using} the variational auto-encoder framework~\cite{kingma2014auto}\ric{). Second,} the model is optimized according to the amortized inference network that learns an efficient mapping from samples to 
\zaiq{variational} distributions~\cite{shu2018amortized}, which is a \ric{method} for efficiently approximating maximum likelihood training. Having inferred the representation vectors of users and items, we can calculate the preference scores of items for each \iadh{user} based on these two types of Gaussian embeddings to make recommendation\iadh{s}.
Our contributions can be summarized as follows~\footnote{The code of our VBCAR	model is publicly available from: https://github.com/mengzaiqiao/VBCAR}:
\begin{enumerate}
\item We propose a variational Bayesian context-aware representation model for grocery recommendation that jointly learns probabilistic user and item representations while the item-user-item triples in \iadh{the} shopping baskets can be \zaiq{reconstructed}. 
\item We use the amortized inference neural network to infer the embeddings of both users and items, which can learn more expressive latent representations by integrating both the non-linearity and Bayesian \iadh{behaviour}.
\item We \iadh{validate} the effectiveness of our proposed model \iadh{using} a real large grocery shopping dataset.
\end{enumerate}

\section{Related Work}

In this section, we briefly discuss two lines of related work, \iadh{namely} methods for grocery recommendation and deep neural network-based methods for recommendation.

\ric{A grocery recommender} is a \iadh{type} of recommender system employed in the domain of grocery shopping to support consumers during their shopping \ric{process}. The most significant difference between the grocery recommendation \iadh{task} and other recommendation \iadh{tasks}, such as video recommendation~\cite{rendle2009bpr} and movie rating prediction~\cite{mnih2008probabilistic}, is that the basket context\iadh{ual} information is more \ric{common and important} in grocery shopping \ric{scenarios}. \iadh{However,} most existing matrix completion-based methods~\cite{rendle2009bpr,mnih2008probabilistic,he2017neural} are unable to incorporate such basket information. Hence, many approaches have been proposed to learn latent representations that incorporate the basket information to enhance the performance of grocery recommendation~\cite{le2017basket,wan2018representing,grbovic2015commerce}, among which Triple2vec~\cite{wan2018representing} is one of the most \ric{effective.} Triple2vec~\cite{wan2018representing} is a recent proposed approach, \iadh{which uses} the Skip-gram model to capture the semantics in \iadh{the} users' grocery basket for product representation and purchase prediction. In this paper, we also apply the Skip-gram model to calculate the likelihood of the basket-based purchase history, but we further \iadh{extend} it to the Bayesian framework \iadh{that represents users and items as Gaussian 
distributions and optimize them} with the Amortized Inference~\cite{kingma2014auto,shu2018amortized}.

Besides the Skip-gram-based models, other deep neural network-based recommendation methods have also achieved success due to the \ric{highly} expressive \ric{nature} of deep learning techniques~\cite{he2017neural,NGCF19,liang2018variational}. For instance, the Neural Collaborative Filtering~\cite{he2017neural} model is a general framework that integrates deep learning into matrix factorization approaches \ric{using} implicit feedback. \iadh{Meanwhile}, Li et al. proposed a collaborative variational auto-encoder ~\cite{li2017collaborative} that learns deep latent representations from content data in an unsupervised manner and also learns implicit relationships between items and users from both content and rating\iadh{s}.  \ric{Additionally, to better capture} context\iadh{ual} information, Manotumruksa et al.~\cite{manotumruksa2018contextual} proposed two gating mechanisms, i.e. a Contextual Attention Gate (CAG) and Time- and Spatial-based \iadh{Gates} (TSG), incorporating both time and geographical information for \ric{(venue) recommendation.} In this work, to further enhance the expressive ability of the learned embeddings for grocery recommendation, we propose to use the variational auto-encoder-based deep neural network~\cite{kingma2014auto} to approximately optimize the variational lower bound.

\section{Methodology}

In this section, we first briefly introduce the basic notations and the problem that we plan to address (\sec{} \ref{sec:notation}). \iadh{Next, we} briefly review the Skip-gram model as well as a state-of-the-art representation model called Triple2vec~\cite{wan2018representing} \iadh{tailored to} grocery recommendation (\sec{} \ref{sec:triple2vec}). \iadh{Then, we} present \iadh{our} proposed representation learning model, i.e. Variational Bayesian Context-Aware Representation (\ours{}), as well as show how to use the learned embeddings for downstream recommendation tasks.

\subsection{Problem Definition and Notations}
\label{sec:notation}
We use $\set{U}=\{u_1,u_2,\cdots,u_N\}$ to denote the set of users and $\set{I}=\{i_1,i_2,\cdots,i_M\}$ to denote the set of items, where $N$ is the number of users and $M$ is the number of items. Then, in \iadh{a} grocery shopping scenario, the users' purchase history can be represented as $\set{S}=\{(u,i,o)\mid u\in \set{U},i\in \set{I},o\in \set{O}\}$ with $\set{O}=\{o_1,o_2,\cdots,o_L\}$ being the set of orders (i.e. baskets). We also use $\mat{Z}^{u}\in \mathbb{R}^{N\times{D}}$ and $\mat{Z}^{i}\in \mathbb{R}^{M\times{D}}$ to denote \iadh{the} latent representation matrices for users and items, respectively, where $D$ denotes the dimension of these latent variables.

Given $\set{U}$, $\set{I}$, $\set{O}$ and $\set{S}$, the task we aim to address in our paper is to infer the latent representation matrices of users and items, i.e. $\mat{Z}^{u}$ and $\mat{Z}^{i}$, so that the missing preference scores of items for each user \ric{that estimate} future \ric{user} purchase probabilities can be predicted \ric{(using} these latent representation matrices\ric{)}.

\subsection{Skip-gram and Triple2vec}
\label{sec:triple2vec}
The Skip-gram model was originally designed for estimating \iadh{word representations} that \iadh{capture} co-occurrence relations between a word and its surrounding words in a sentence~\cite{mikolov2013distributed}. It aims \iadh{to maximize} the \ric{log-likelihood} of a target entity (word) $v$ predict\ric{ing} contextual entities (words)  $C_v$:
\begin{equation}
\log p(C_{v} \mid v)= \sum_{v^{\prime} \in C_{v}} \log P\left(v^{\prime} | v\right),
\end{equation}
where $P\left(v^{\prime}\mid v\right)$ is defined by the softmax formulation $P\left(v^{\prime} | v\right)=\frac{\exp \left(f_{v}^{T} g_{v^{\prime}}\right)}{\sum_{v^{\prime \prime}} \exp \left(f_{v}^{T} g_{v^{\prime \prime}}\right)}$ with $f_v$ and $g_{v^{'}}$ being the latent representations of the target entity and its contextual entities\iadh{,} respectively.

\begin{figure*}
	\begin{centering}
		\includegraphics[width=10cm]{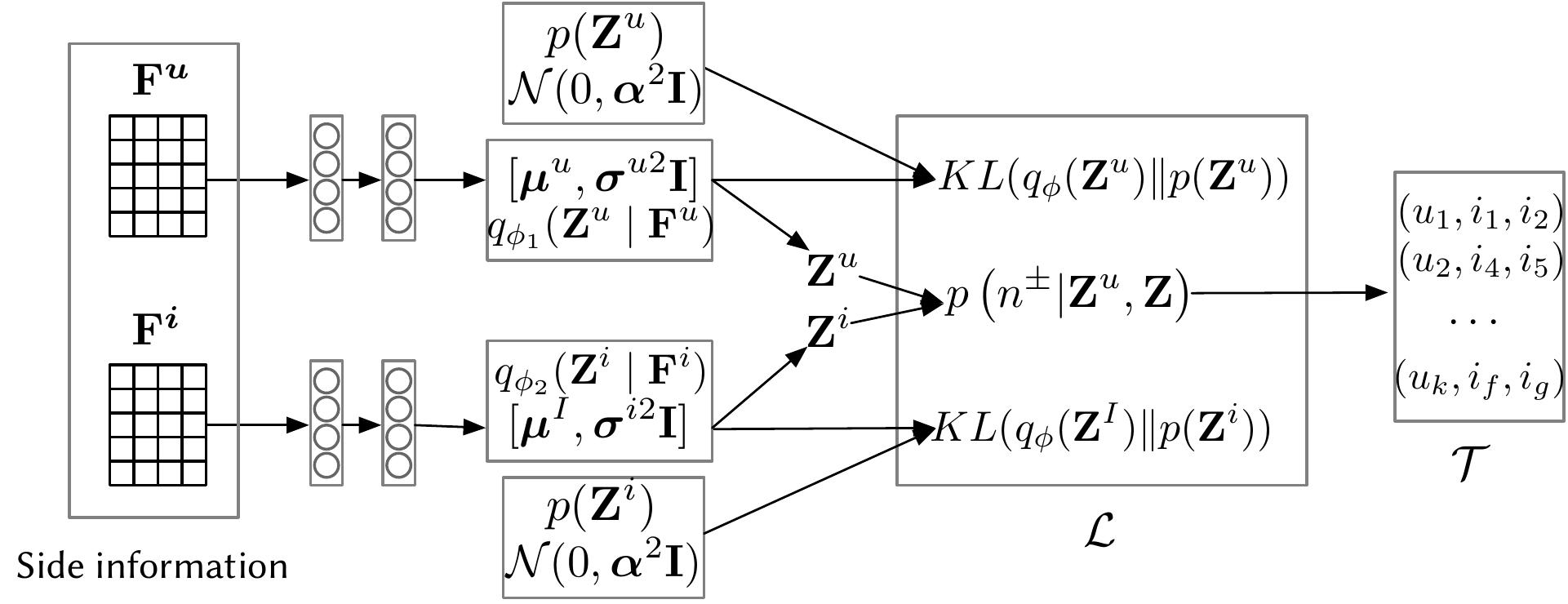}
		\par\end{centering}	
	\caption{\label{fig:overview}The architecture of our proposed \ours{} model. The model takes the user and item one-hot identity representation, i.e. $\mat{F}^u$ and $\mat{F}^i$, as input and outputs Gaussian distributions with means and variances as latent embeddings for all users and items. The model then uses the deterministic variables $\mat{Z}^u$ and $\mat{Z}^i$, reparameterized from their Gaussian distributions, to predict the sampled triples.}
	\vspace{-1em}
\end{figure*}

The Triple2vec~\cite{wan2018representing} model further extends the Skip-gram model for capturing  \iadh{co-purchase product} relationships within users\iadh{'} baskets according to sampled triples from \iadh{the} grocery shopping data\ric{. Here} each triple reflects two items purchased by the same user in the same basket. Specifically, Triple2vec samples a set of triples $\mathcal{T}=\{(u,i,j)\mid (u,i,o)\in \set{S}, (u,j,o)\in\set{S} \}$ from the purchase history $\set{S}$ as the purchase context for training and assumes that a triple \zaiqm{$(u,i,j) \in \mathcal{T}$} is generated by \iadh{a} probability $\sigma$ calculated by the function of  \zaiqm{$p((u, i, j)\mid \mat{z}^{u}_u,\mat{z}^{i}_i,\mat{z}^{i}_j)$}:
\begin{align}
\label{eq:likelihood}
\sigma = \zaiqm{p((u, i, j)\mid \mat{z}^{u}_u,\mat{z}^{i}_i,\mat{z}^{i}_j)}=P(i | j, u)P(j | i, u)P(u | i, j),
\end{align}
where $\mat{z}^{u}_u\in \mat{Z}^{u}$ and $\mat{z}^{i}_i, \mat{z}^{i}_j \in \mat{Z}^{i}$ are the latent representations of user  $u$ and items $i$ and $j$, respectively, $P(i\mid j, u)=\frac{\exp \left(\mat{z}^{iT}_{i} (\mat{z}^{i}_{j}+\mat{z}^{u}_{u})\right)}{\sum_{i^{\prime}} \exp \left(\mat{z}^{iT}_{i^{\prime}} (\mat{z}^{i}_{j}+\mat{z}^{u}_{u})\right)}$ and $P(u \mid i, j)=\frac{\exp \left(\mat{z}^{uT}_{u} (\mat{z}^{i}_{i}+\mat{z}^{i}_{j})\right)}{\sum_{u^{\prime}} \exp \left(\mat{z}^{uT}_{u^{\prime}} (\mat{z}^{i}_{i}+\mat{z}^{i}_{j})\right)}$.
The Skip-gram based models ~\cite{wan2018representing,grbovic2015commerce} can learn representations for users and items at scale, and with the aid of basket information they have \ric{previously been shown to be effective} for grocery recommendation. However, these models represent each user and item by single deterministic points in a low-dimensional continuous space, which limits the expressive ability of their embeddings and recommendation performance. To address \iadh{this} problem, we propose a new Bayesian Skip-gram model that represents users and items by Gaussian distributions, as illustrated in \sec{} \ref{sec:baysskpgram}. \iadh{Then,} we describe how to approximately optimize the Bayesian Skip-gram model with \iadh{a} Variational Auto-encoder and the amortized Inference (\sec{} \ref{sec:inference}). \ric{We provide an overview of our overall proposed model in} \fig{} \ref{fig:overview}.

\subsection{The Variational Bayesian Context-aware Representation Model}
\setlength{\abovedisplayskip}{5pt}
\setlength{\belowdisplayskip}{5pt}

\subsubsection{Bayesian Skip-gram Model}
\label{sec:baysskpgram}
Here we present our proposed Variational Bayesian Context-aware Representation model, i.e. \ours{}, \iadh{which} represents the users and items  as random variables, i.e. $\mat{Z}^{u}$ and $\mat{Z}^{i}$ \iadh{that} are independently generated according to their priors.  Like other probabilistic methods for embedding~\cite{meng2019co} and recommender systems~\cite{he2017neural,liang2018variational}, these priors are assumed to be the standard Gaussian distributions:
\begin{align}
p\left(\mat{Z}^{u}\right)=&\mathcal{N}\left(0, \alpha^{2} \mat{I}\right), &
p\left(\mat{Z}^{i}\right)=&\mathcal{N}\left(0, \alpha^{2} \mat{I}\right)
\end{align}
where $\alpha^{2}$ is \iadh{the} same hyperparameter for all the priors - we used the default setting of $\alpha=1$ in our paper, following~\cite{kingma2014auto}.

\ric{Consider past purchase} triples $(u,i,j)\in\set{T}$ that are sampled \ric{from historical} grocery shopping data~\cite{wan2018representing}. These sampled triples are \textit{positive examples} that should be precisely predicted according to the latent variables of users and items. We use $n+$ to denote the number of times that a given \iadh{triple} is observed in the \textit{total sample} $\set{T}$.
 Then $n+$ is a \textit{sufficient statistic} of the Skip-gram model, and it \iadh{contributes} to the likelihood $p(n+\mid \mat{z}^{u}_u,\mat{z}^{i}_i,\mat{z}^{i}_j)=\sigma^{n+}$~\cite{barkan2017bayesian}.
Thus, one needs to \ric{also} construct \ric{an associated} rejected triples set (i.e. $n-$, \textit{negative examples}) that are not in the total sample so that we can conduct \iadh{an} efficient \textit{negative sampling} for approximate optimization~\cite{mikolov2013distributed}. We let $n{\pm}=\{n+, n-\}$ be the combination of both positive and negative examples, then the likelihood of this complete purchase context is obtained by:
\begin{equation}
\log p\left(n{\pm} | \mat{Z}^{u}, \mat{Z}^{i}\right)=\sum_{(v_i, v_j, u_u) \in \mathcal{T}}\log\sigma + \sum_{(v_i, v_j, u_u) \notin \mathcal{T}}\log(1-\sigma),
\end{equation}
where $\sigma$ is calculated by the same function as in Triple2vec~\cite{wan2018representing} (i.e.\ $p((u, i, j)\mid \mat{z}^{u}_u,\mat{z}^{i}_i,\mat{z}^{i}_j)$ in \eq{} (\ref{eq:likelihood})).
\subsubsection{The Variational Evidence Lower Bound and Amortized Inference}
\label{sec:inference}
Since we assume that both $\mat{Z}^{u}$ and $\mat{Z}^{i}$ are random variables, \iadh{the} exact inference of their posterior density is intractable due to \iadh{the non-differentiable marginal likelihood} $p\left(n^{ \pm}\right)$~\cite{kingma2014auto}. Variational Bayes resolves this issue by constructing a tractable lower bound of the logarithm marginal likelihood and maximizing the lower bound instead~\cite{blei2017variational}. Following the Variational Autoencoding framework~\cite{kingma2014auto}, we also solve \iadh{this} problem by \iadh{introducing} the two variational distributions to formulate a tractable lower bound and \iadh{optimize} the lower bound by the Amortized Inference~\cite{shu2018amortized}. To infer the users' and items' embedding, we start by formulating the logarithm marginal likelihood of $n^{ \pm}$:
\begin{align} 
\label{eq:elbo}
\log p\left(n^{ \pm}\right)=&\log\mathbb{E}_{q_{\phi}\left(\mat{Z}^{u},\mat{Z}^{i}\right)}\left[\frac{p\left(n^{ \pm},\mat{Z}^{u}, \mat{Z}^{i}\right)}{q_{\phi}\left(\mat{Z}^{u}, \mat{Z}^{i}\right)}\right]
\end{align}
\begin{align} 
\nonumber \ge&\mathbb{E}_{q_{\phi}\left(\mat{Z}^{u}, \mat{Z}^{i}\right)}\left[\log\frac{p\left(n^{ \pm},\mat{Z}^{u}, \mat{Z}^{i}\right)}{q_{\phi}\left(\mat{Z}^{u}, \mat{Z}^{i}\right)}\right]\\\nonumber
=&\mathbb{E}_{q_{\phi}\left(\mat{Z}^{u}, \mat{Z}^{i}\right)}\left[\log p\left(n^{ \pm}\mid \mat{Z}^{u}, \mat{Z}^{i}\right)\right]\\\nonumber
&-KL\left(q_{\phi}\left(\mat{Z}^{u}, \mat{Z}^{i}\right)\|p(\mat{Z}^{u}, \mat{Z}^{i})\right)\\\nonumber
\overset{\underset{\mathrm{def}}{}}{=}&\set{L},
\end{align}
where the inequation of the second line is derived from the Jensen's inequality\iadh{;} $\set{L}$ is called the Evidence Lower BOund (ELBO) of the observed triple context~\cite{kingma2014auto}\iadh{;}
$KL(\cdot\|\cdot)$ is the Kullback-Leibler (KL) divergence and $q_{\phi}(\mat{Z}^{u}, \mat{Z}^{i})$ is the variational distribution, which can be factorized in a mean-field form:
\begin{align}
q_{\phi}(\mat{Z}^{u},\mat{Z}^{i})=q_{\phi_{1}}(\mat{Z}^{u})q_{\phi_{1}}(\mat{Z}^{i}),
\end{align}
where $\phi_{1}$ and $\phi_{2}$ are the trainable parameters of the inference models (encoders).
In order to get more expressive latent factors of users and items, we consider that the variational distributions $q_{\phi_1}(\mat{Z}^{u})$ and $q_{\phi_2}(\mat{Z}^{i})$ are Gaussian distributions and are encoded from the identity codes of users and items such that we have:
\begin{align}
q_{\phi_{1}}\left(\mat{Z}^{u}\mid \mat{F}^{u}\right)&=\mathcal{N}\left({\boldsymbol{\mu}}^{u}, \boldsymbol{\sigma}^{u2} \mathbf{I}\right),\\
q_{\phi_{2}}\left(\mat{Z}^{i}\mid \mat{F}^{i}\right)&=\mathcal{N}\left({\boldsymbol{\mu}}^{i}, \boldsymbol{\sigma}^{i2} \mathbf{I}\right),
\end{align}
where $\mat{F}^{u}\in \mathbb{R}^{N\times{F_1}}$ and $\mat{F}^{i}\in \mathbb{R}^{M\times{F_2}}$ are \iadh{the} identity representation (can be one-hot or binary encoded) of users and items respectively, with $F_1$ and $F_2$ being the dimension of their identity representation respectively, and $\boldsymbol{\mu}^{u}$, $\boldsymbol{\sigma}^{u2}$, $\boldsymbol{\mu}^{i}$ and $\boldsymbol{\sigma}^{i2}$ are inferred by the encoder networks. Specifically, the parameters of these Gaussian embeddings are encoded from their identity codes, i.e.\ $\mat{F}^{u}$ and $\mat{F}^{i}$, according to two two-layer fully-connected neural networks:
\begin{align}
\label{eq:inference1}
[\boldsymbol{\mu}^{u}, \boldsymbol{\sigma}^{u2} \mathbf{I}]=&\mat{W}^{u}_2\textbf{tanh}(\mat{W}^{u}_1\mat{F}^{u}+\mat{b}^{u}_1)+\mat{b}^{u}_2,\\
\label{eq:inference2}
[\boldsymbol{\mu}^{i}, \boldsymbol{\sigma}^{i2} \mathbf{I}]=&\mat{W}^{i}_2\textbf{tanh}(\mat{W}^{i}_2\mat{F}^{i}+\mat{b}^{i}_2)+\mat{b}^{i}_2,
\end{align}
\looseness -1 where \textbf{tanh} is the non-linearity activation function, \iadh{and} $\mat{W}^{u}_1$, $\mat{W}^{u}_2$, $\mat{b}^{u}_1$, $\mat{b}^{u}_2$, $\mat{W}^{i}_1$, $\mat{W}^{i}_2$, $\mat{b}^{i}_1$ \& $\mat{b}^{i}_2$ are trainable parameters of the neural networks.

Since we assume the priors and the variational posteriors are Gaussian distributions, the KL-divergence terms in \eq{} \iadh{(\ref{eq:elbo})} have analytical forms. By using the Stochastic Gradient Variational Bayes (SGVB) estimator and the reparameterization trick~\cite{kingma2014auto}, we can directly optimize \iadh{ELBO} by sampling deterministic and differentiable embedding samples from the \iadh{inferred variational} distributions:
\begin{align}
\mat{Z}^{u}&=\boldsymbol{\mu}^{u}+\boldsymbol{\sigma}^{u2} \odot \boldsymbol{\epsilon}^{(l)}, \boldsymbol{\epsilon}^{(l)}\sim \mathcal{N}\left(0, \mat{I}\right),\\\nonumber
\mat{Z}^{i}&=\boldsymbol{\mu}^{i}+\boldsymbol{\sigma}^{i2} \odot \boldsymbol{\epsilon}^{(l)}, \boldsymbol{\epsilon}^{(l)}\sim \mathcal{N}\left(0, \mat{I}\right),
\end{align}
 to approximate and regularize maximum likelihood training, which is also \iadh{referred to as} \textbf{amortized inference}~\cite{shu2018amortized}.

\subsection{Recommendation Tasks}
\label{sec:reco_task}
Our model infers the embeddings of both users and items according to the variational auto-encoder and \iadh{represents} them by means of their variational Gaussian distributions. Since we have taken advantage of the basket information, having obtain\iadh{ed} the embedding of users and items \ric{in} our \ours{} \ric{model}, we can follow \iadh{a} similar \iadh{approach to}~\cite{wan2018representing} \ric{tackling both} next-basket product recommendation and within-basket product recommendation\ric{:}
\begin{enumerate}
	\item Next-basket product recommendation: Recommending a given user $u$ with products for the next basket, we can obtain a preference score $s_{ui}=\textbf{dot} \left(\mat{z}_u^{u},\mat{z}_i^{i}\right)$\footnote{\textbf{dot} is the dot product for two vectors.} for each item $i$, then return the top-$K$ items with the highest preference scores.
	\item Within-basket product recommendation: If \iadh{the} products in the current basket $b$ are given, we can first compute a preference score of item $i$ for user $u$ by: $s_{ui}=\textbf{dot}(\mat{z}_u^{u}+\sum_{i^{\prime}\in b}\mat{z}_i^{i^{\prime}},\mat{z}_i^{i})$, then return the top-$K$ preference \iadh{score} items as recommendation\iadh{s}.
\end{enumerate}
In this paper, we only evaluate the performance of our \iadh{model} based on the next-basket product recommendation and leave the evaluation of within-basket product recommendation for future work.

\section{Experiments}

\iadh{In the following, we first introduce the research questions we aim to answer in this paper (Section~\ref{rqs}). Next, we describe our experimental setup (Section~\ref{setup}), followed by our results and analysis (Section~\ref{results})}.

\subsection{Research \iadh{Q}uestions} \label{rqs}

\iadh{In this paper, we aim to answer the following two research questions:}
\begin{enumerate}[(\textbf{RQ}1)]
	\item Can our \iadh{proposed} model outperform \iadh{the} Triple2vec model for grocery recommendation?
	\item Can our Bayesian model learn more expressive representations of users and items than Triple2vec?
\end{enumerate}
\subsection{Experimental \iadh{Setup}} \label{setup}
\noindent\textbf{Dataset.}
We evaluate our model \iadh{using} the \textbf{Instacart}~\cite{wan2018representing} dataset, which is a public \iadh{large} grocery shopping \iadh{dataset} from the Instacart Online Grocery Shopping Website\footnote{\url{https://www.instacart.com/datasets/grocery-shopping-2017}}. This dataset contains over 3 million grocery orders and 33.8 million interactions from 0.2 million users and 50 thousand items. We first clean the dataset by filtering users and items \iadh{using a number of thresholds}. \ric{In particular,} users that have less than 7 orders or less than 30 items\ric{, as well as} items that were purchased by less than 16 users in the purchase history were removed. \iadh{Next, we} uniformly sample different percentages of users and items to construct different \iadh{sizes} of evaluation dataset. For model evaluation, we split all the sampled datasets into training (80\%) and testing (20\%) sets according to the temporal order of baskets. \tab{} \ref{tab:dataset} shows the statistics of \ric{these} datasets.

\begin{table}[H]
	\caption{Statistics of the datasets in \iadh{used in our} experiments.}
	\vspace{-1em}
	\label{tab:dataset}
	\begin{tabular}{@{}l*{4}c@{}}
		\toprule 
		\textbf{Percentage} & \textbf{\#Orders} & \textbf{\#Users} & \textbf{\#Items} & \textbf{\#Interactions}
		\tabularnewline 
		\cmidrule{2-5}
		\textbf{$5\%$} & 47,207 & 5,679 & 1,441 & 354,946
		\tabularnewline
		\textbf{$10\%$} & 154,285 & 11,888 & 3,124 & 1,103,361
		\tabularnewline
		\textbf{$25\%$} & 527,431 & 46,850 & 9,174 & 4,010,904
		\tabularnewline
		\textbf{$50\%$} & 1,186,957 & 59,549 & 16,121 & 12,217,555
		\tabularnewline 
		\textbf{$100\%$} & 2,741,332 & 119,098 & 32,243 & 29,598,689
		\tabularnewline 
		\bottomrule
	\end{tabular}
	\vspace{-0.5em}
\end{table}
\noindent\textbf{Baseline and Evaluation Metrics.}
\ric{To provide a} fair comparison, \ric{we use} Triple2vec~\cite{wan2018representing} as \ric{a state-of-the-art} baseline\ric{, since} Triple2vec \ric{incorporates} basket information \ric{in a similar way to our proposed VBCAR model.} 
We evaluate the effectiveness of our model \ric{for next-basket (grocery) product recommendation}, where we evaluate the top-K items recommended by each model. We \ric{report the standard recommendation evaluation metrics Recall@K and NDCG@K}~\cite{NGCF19,he2017neural,wan2018representing} to \iadh{evaluate} the preference ranking performance. \ric{We report} result\iadh{s} \iadh{for} $K=10$ in our subsequent experiments\ric{, however we observed similar results when testing other} values of $K$ (e.g. 5 and 20)\ric{.}

\subsection{Result\iadh{s} and Analysis} \label{results}

\begin{table}[htp]
	\caption{\zaiq{Overall performance on item recommendation. The best performing result is highlighted in bold; and ${}^*$ denotes a significant difference compared to the baseline result, according to the paired t-test p < 0.01.}}
	\label{tab:over_perform}
	\vspace{-1em}
	\begin{tabular}{lcccc} 
		\toprule
		\multirow{2}{*}{\textbf{Dataset}}  &    \multicolumn{2}{c}{\textbf{Triple2vec}}  &  \multicolumn{2}{c}{\textbf{VBCAR}} \\
		\cmidrule(l){2-3} \cmidrule(l){4-5}
		& NDCG@10 & Recall@10 & NDCG@10 & Recall@10 \\
		\cmidrule{1-5}
		\textbf{5\%} & \zaiq{0.557} & \zaiq{0.708} & \textbf{\zaiq{0.731${}^*$}} & \textbf{\zaiq{0.748${}^*$}} \\ 
		\textbf{10\%} & \zaiq{0.558} & \zaiq{0.664} & \textbf{\zaiq{0.723${}^*$}} & \textbf{\zaiq{0.720${}^*$}} \\
		\textbf{25\%} & \zaiq{0.626} & \zaiq{0.608} & \textbf{\zaiq{0.686${}^*$}} & \textbf{\zaiq{0.628${}^*$}} \\
		\textbf{50\%} & 0.708 & 0.525 & \zaiqm{\textbf{0.719${}^*$}} & \zaiqm{\textbf{0.643${}^*$} }\\
		\textbf{100\%} & 0.726 & 0.660 & \zaiqm{\textbf{0.768${}^*$}} & \zaiqm{\textbf{0.742${}^*$}} \\
		\bottomrule
	\end{tabular}
	\vspace{-1em}
\end{table}

To answer \textbf{RQ1}, we evaluate our model as well as the triple2vec \iadh{baseline} on the task of item recommendation with \iadh{the} same size of triple samples (i.e.\ 1 million). \tab{} \ref{tab:over_perform} shows the overall performance of our proposed model as well as \iadh{that of} the baseline method. For both the Triple2vec and our proposed VBCAR approach, we empirically set \iadh{the} embedding size to be 64 and train \iadh{both} models, with a batch size of 512 and a RMSprop optimizer. From \tab{} \ref{tab:over_perform}, we can clearly see that our \ours{} model performs better than Triple2vec on all the datasets. This result suggests that our model can learn more expressive latent representations by integrating both non-linearity and \iadh{a} Bayesian \iadh{behaviour}.

\begin{figure}[t!]
	\centering
	\subfigure[NDCG@10 performance by different triple size]{\includegraphics[width = 2.2in]{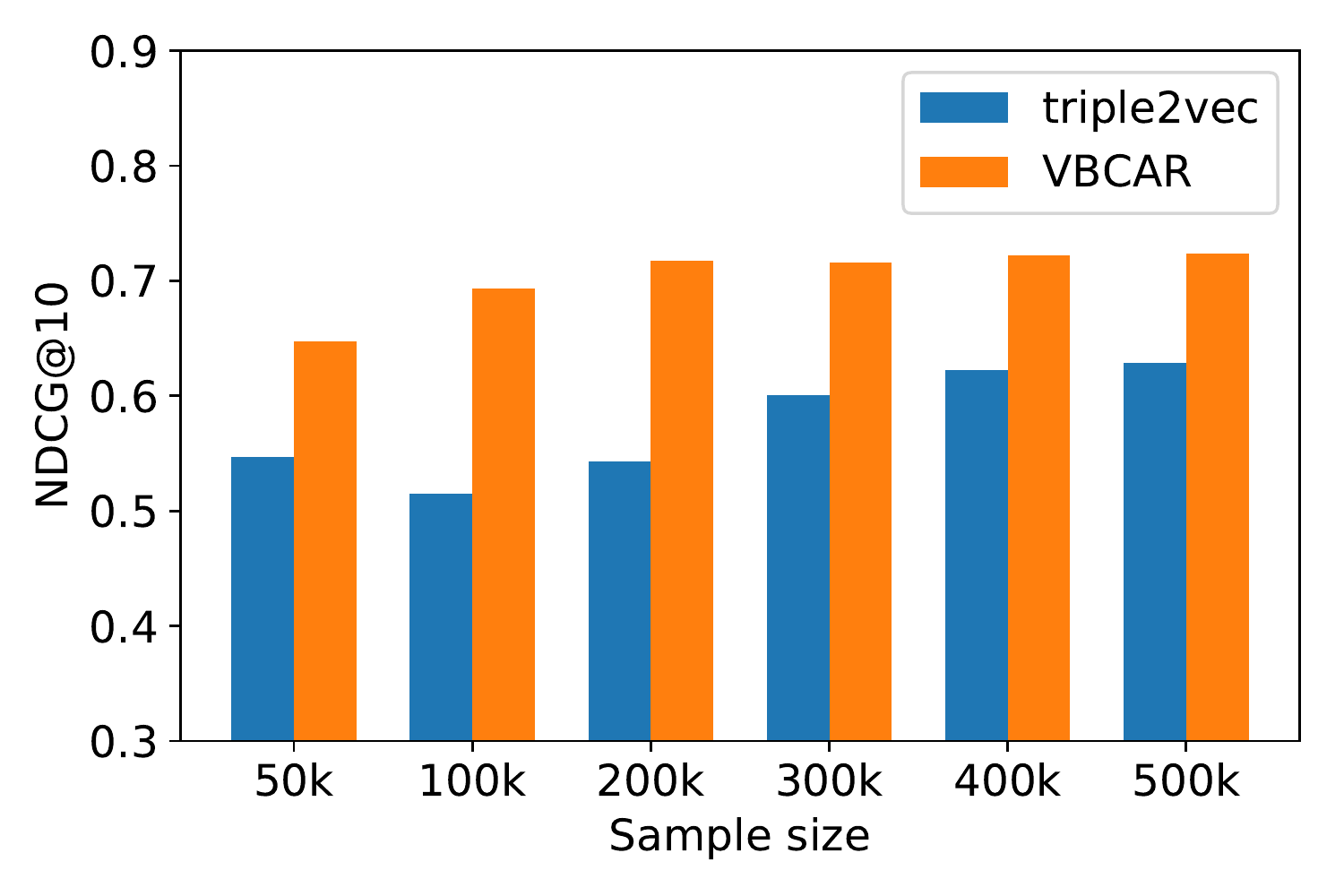}}\\
	\subfigure[Recall@10 performance by different triple size]{\includegraphics[width = 2.2in]{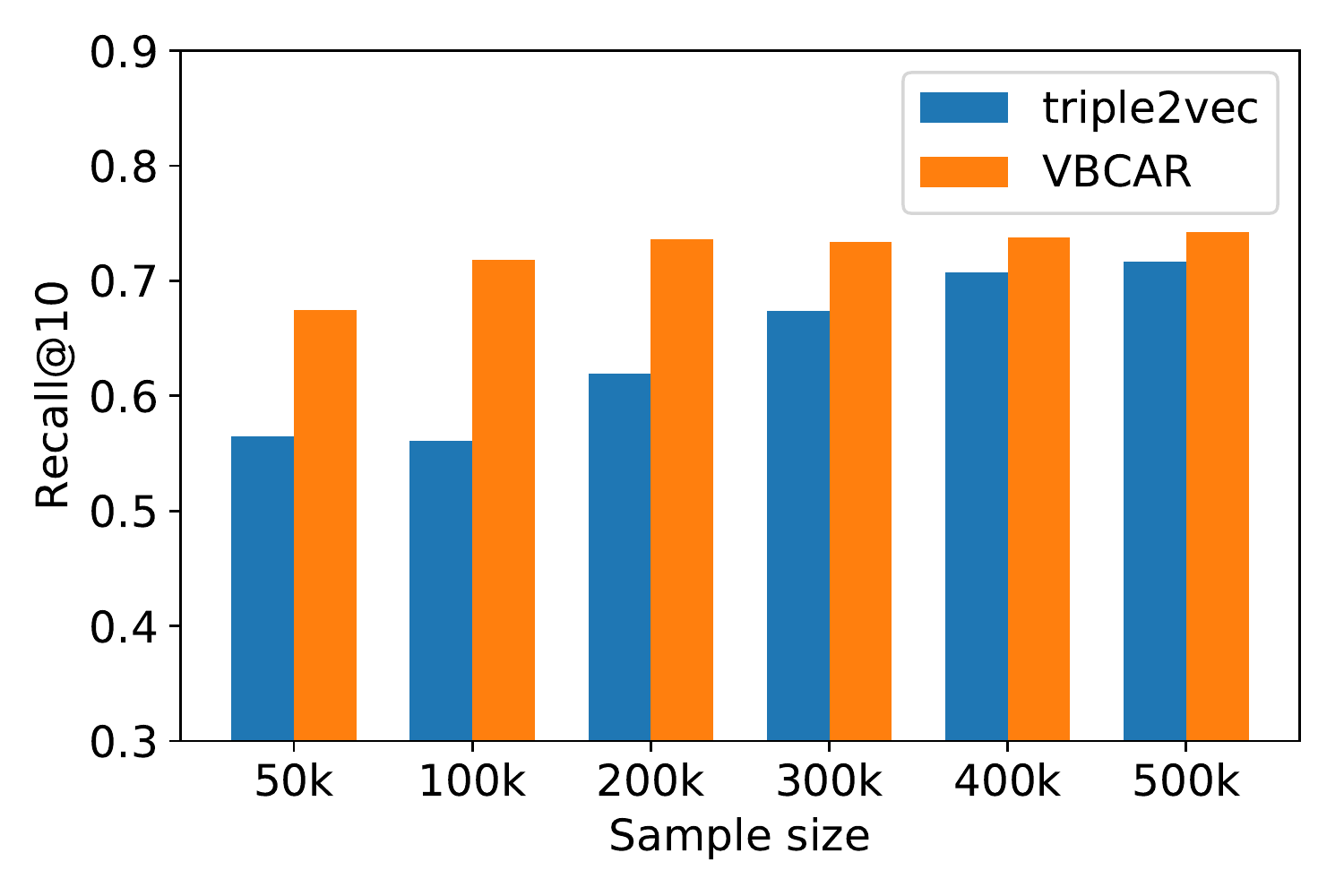}}
	\vspace{-1em}
	\caption{\label{fig:size_comp}Performance comparison for the various sample sizes on 5\% of the Instacart data.}
\end{figure}


To further validate this argument (\textbf{RQ2}), we also compare the recommendation performance of our \ours{} \iadh{model} \iadh{with} Triple2vec \iadh{using a} different number of triple sample\iadh{s}. \fig{} \ref{fig:size_comp} shows the NDCG@10 and Recall@10 \iadh{performances} \iadh{for} triple sample size\iadh{s} ranging from 50k to 500k. In \iadh{these experiments}, we set the embedding dimension \iadh{to} 64, \iadh{while the} other parameters \ric{for both models} are tuned to be optimal except \iadh{the} \ric{fixed} triple sample size. \ric{Again, w}e can clear\iadh{ly} observe that our \ours{} \iadh{model} outperforms Triple2vec in \iadh{terms of both metrics}  and \iadh{on all} triple sample size\iadh{s}. \iadh{Moreover,} the gap between the performance of \ours{} \iadh{model} and Triple\iadh{2vec} is larger \iadh{on small sample sizes}. This result validates \iadh{our hypothesis} that our \ours{} model can learn more expressive latent representations with limited input samples.

\section{Conclusions}
In this paper, we have proposed the VBCAR \iadh{model}, a variational Bayesian context-aware representation model for grocery recommendation. Our model was built based on the variational Bayesian Skip-gram framework coupled with the amortized inference. Experimental result\iadh{s} on \iadh{the} Instacart dataset \iadh{show} that our VBCAR \iadh{model} can learn more expressive representations of users and items than Triple2vec and \iadh{does} significantly outperform Triple2vec \ric{under} \iadh{both the} NDCG and Recall \iadh{metrics}. Indeed, we observe up to a $31\%$ increase in recommendation effectiveness over Triple2vec (under NDCG@10). For future work, we plan to extend our model to infer latent representations for new users and new items by taking the side information \iadh{about} users and items into account. 
\vspace{-0.5em}
\subsection*{Acknowledgements}
\zaiqm{The research leading to these results has received funding from the European Community's Horizon 2020 research and innovation programme under grant agreement $\text{n}^{\circ}$ 779747.}
\vspace{-0.5em}
\bibliographystyle{ACM-Reference-Format} 
\bibliography{bib}
\end{document}